\documentclass[twocolumn,twoside,slac_two]{revtex4}
\usepackage{graphicx}
\usepackage{fancyhdr}
\begin{document}

\title{The Physics of Charm: Recent Experimental Results}
\author{Jim Napolitano}
\affiliation{Rensselaer Polytechnic Institute, Troy, NY 12180 USA}

\begin{abstract}
We review the most recent results from experiments studying systems containing charmed quarks. The selection reflects the presenter's bias, and there is an emphasis on decays of open charm. We discuss precision measurements of various sorts, various new states in the charmonium system, measurements aimed at testing Lattice QCD, and the latest searches for charm mixing. We conclude with a discussion of upcoming experiments at existing and future facilities
\end{abstract}

\maketitle

\thispagestyle{fancy}

\section{INTRODUCTION}

Charm is a unique laboratory for studying the fundamental interactions. Its mass is much larger than $\Lambda_{\rm QCD}$, so it qualifies as a ``heavy quark'' as it is amenable to calculations based on effective heavy quark theory and the nonrelativistic quark model. These also make it suitable as a testing ground for recent calculations using unquenched Lattice QCD.

On the other hand, hadrons composed of charm quarks are light enough so that most of their decay modes can be enumerated. Large fractions of the decay modes of $D$ mesons, for example, have been determined rather precisely.

Furthermore, unlike $s$ and $b$ quarks, CKM-allowed decays are accessible to charm quarks. This leads, among other things, to the suppression of $CP$ violation and $D^0$ mixing within the Standard Model. In other words, decays of charmed hadrons can be ideal laboratories for searching for ``new physics."

The presenter is a member of the CLEO-c collaboration. This is an incarnation of CLEO where the detector and CESR accelerator are optimized for the production and study of charm. This review, however, incorporates results from several other experiments as well, including BaBar, Belle, BES~II in $e^+e^-$ annihilation, photoproduction from FOCUS, and the D0, CDF, and E835 experiments in hadroproduction.

Nevertheless, the selection of results presented here reflect the presenter's bias. We group results in four categories:
\begin{itemize}
\item ``Precision'' measurements. We discuss results with small error bars, producing stringent limits, or tying up old loose ends.
\item New states. In particular, we review some recent results in the charmonium system. {\em Note: See also the talk by R.\ Waldi in these proceedings.}
\item Confronting Lattice QCD. We compare some measurements to recent high precision ($\sim1\%$) calculations made on the lattice.
\item $D^0$ mixing and tests of $CP$ violation. Various results have recently been released, and novel measurements are on the way.
\end{itemize}

\section{PRECISION MEASUREMENTS}

\subsection{The $D_s$ Lifetime from FOCUS}
\begin{figure}[ht]
\begin{center}
\includegraphics[angle=-90,width=80mm]{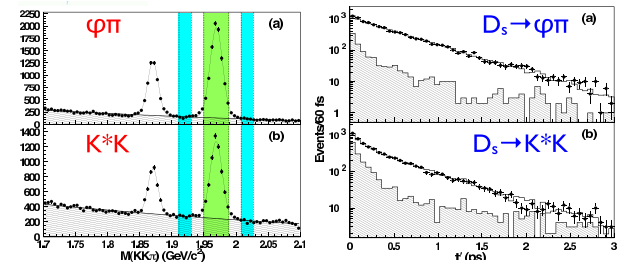}
\caption{Lifetime of the $D_s^\pm$ from FOCUS. The left side shows the signal and background regions, and also displays the Cabibbo-suppressed $D^\pm$ signal. The right shows the proper time distributions for signal and background events.\label{fig:DsLifetime}}
\end{center}
\end{figure}

The FOCUS collaboration at FermiLab has determined the lifetime of the $D_s^\pm$ using high energy, fixed target photoproduction~\cite{Link:2005ew}. They produce $D_s$ using the reaction $\gamma p\to(K^+K^-\pi^\pm)X$, enhancing signal to background by identifying the detached decay vertex. See Fig.~\ref{fig:DsLifetime}.
In each case, sidebands are used to determine the background event distributions, and care is taken to exclude Cabibbo-suppressed $D^\pm$ events.
The analysis is done separately for $\phi\pi$ and $K^*K$ final states.  The results are consistent between both modes, and they determine a lifetime $\tau(D_s)=507.4\pm5.5\pm5.1$~fs

\subsection{Hadronic Decays of $D_s$}
\label{sec:DsHad}
\begin{figure}[ht]
\begin{center}
\includegraphics[width=80mm]{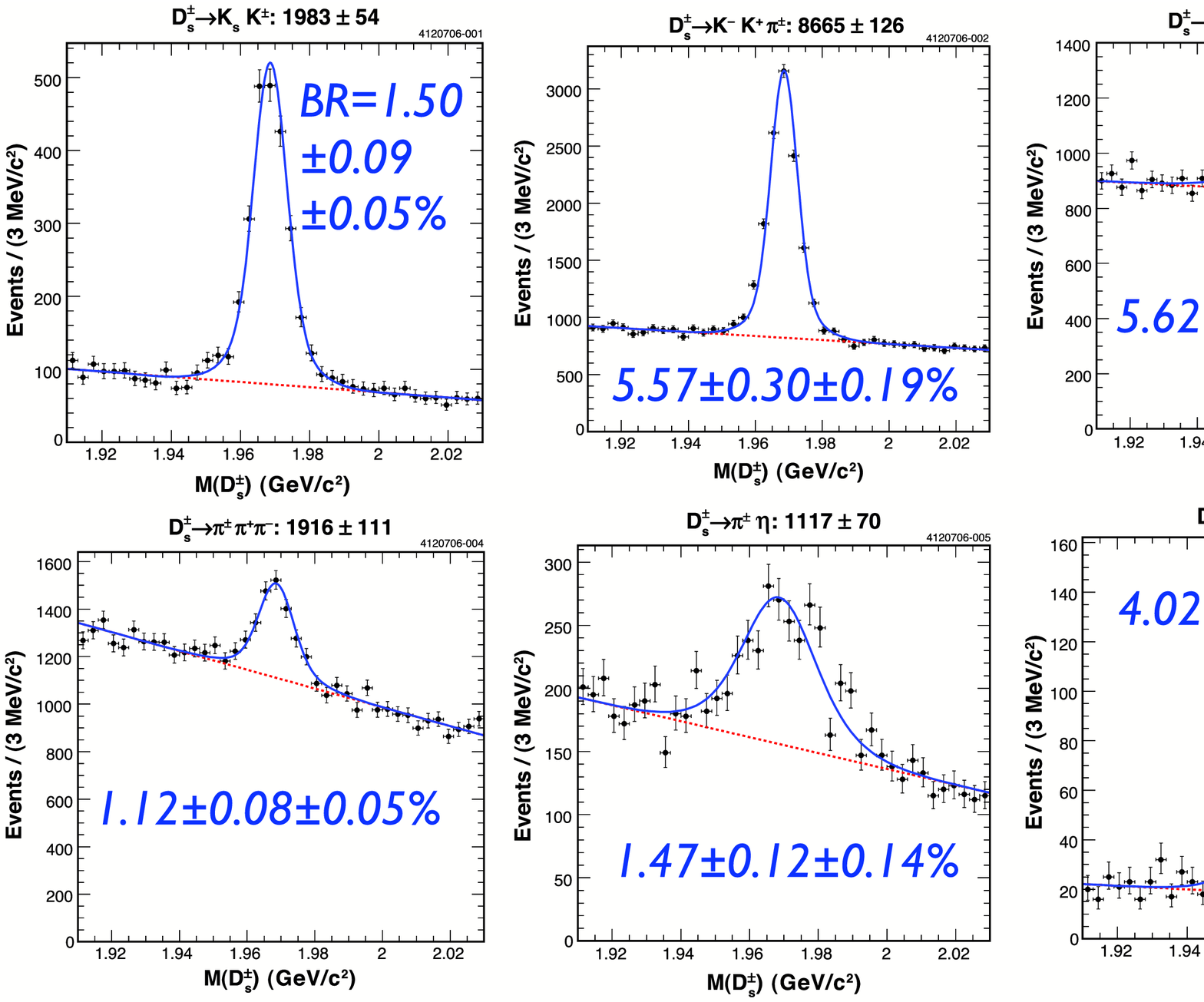}
\caption{Hadronic branching fraction signal and values for various $D_s^\pm$ hadronic decay modes from CLEO-c. Decay modes are (left to right, top) $K_SK^\pm$, $K^-K^+\pi^\pm$, $K^-K^+\pi^\pm\pi^0$, (bottom) $\pi^\pm\pi^+\pi^-$, $\pi^\pm\eta$, and $\pi^\pm\eta^\prime$.\label{fig:DsHad}}
\end{center}
\end{figure}

Using the reaction $e^+e^-\to D_sD_s^*$, CLEO-c has made precise measurements of various hadronic branching fractions for $D_s$ mesons~\cite{Adam:2006me}. These results are from the initial running period at $E_{\rm cm}=4.17$~GeV, yielding 195~fb$^{-1}$ of integrated luminosity. Backgrounds are rather low, and statistical and systematic errors are comparable to each other in all cases. Invariant mass plots and preliminary branching ratio values are shown in Fig.~\ref{fig:DsHad}.

\subsection{Search for $D^+\to\pi^+\mu^+\mu^-$ from D0}
Enormous charm samples are available in principle from high energy $\bar pp$ collisions. The D0 experiment at FermiLab has carried out a search for ``new physics'' in the rare decay $D^+\to\pi^+\mu^+\mu^-$~\cite{D0PiMuMu}. This is a particularly interesting example, because there is an appreciable branching fraction for $D^+\to\pi^+\phi$ (and also $D_s^+\to\pi^+\phi$, as seen in Fig.~\ref{fig:DsLifetime}) and a small, but observable, branching ratio for $\phi\to\mu^+\mu^-$.

\begin{figure}[ht]
\begin{center}
\includegraphics[width=40mm]{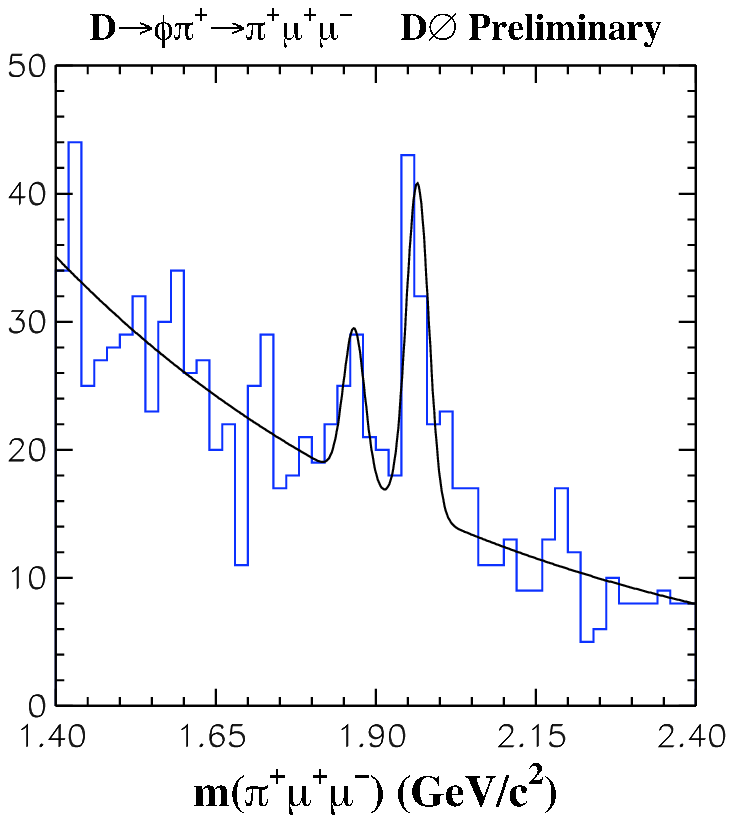}\hfill
\includegraphics[width=40mm]{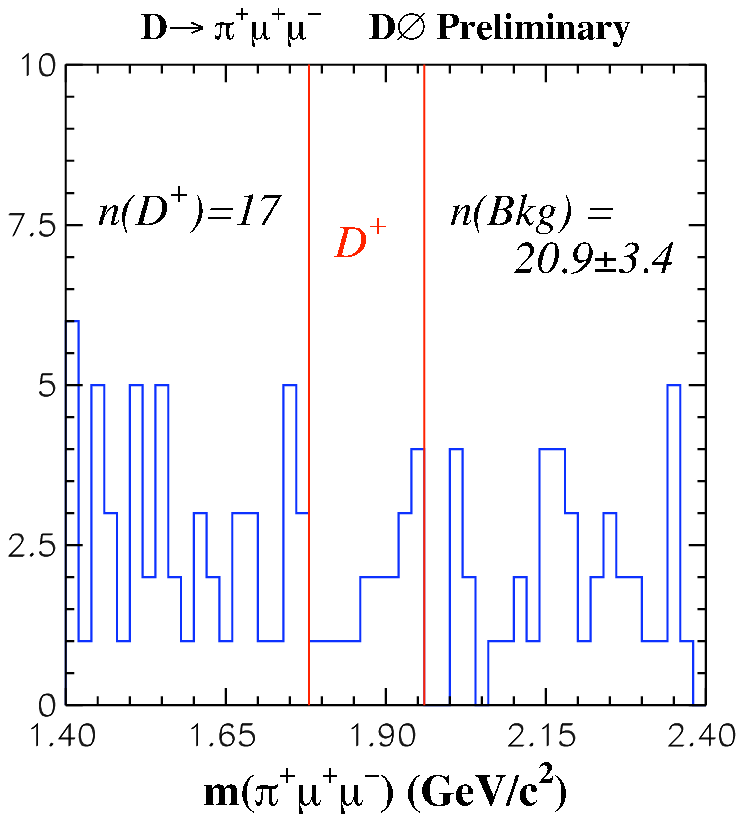}
\caption{Data from D0 at FermiLab, for $D^+\to\pi^+\phi\to\pi^+\mu^+\mu^-$ (left) and direct $D^+\to\pi^+\mu^+\mu^-$ (right).\label{fig:PiMuMu}}
\end{center}
\end{figure}

Figure~\ref{fig:PiMuMu} shows the results of the search.
The $\pi^+\mu^+\mu^-$ mass is plotted, for $\mu^+\mu^-$ mass inside (left) and outside (right) the $\phi$ window. Clear signals are seen for $D^+\to\pi^+\phi$ and $D_s^+\to\pi^+\phi$, but there is no evidence for a direct $D^+\to\pi^+\mu^+\mu^-$ decay, An upper limit on the branching ratio is set at $4.7\times10^{-6}$.

\subsection{Decays of the $\psi(3770)$}
Since the early days of charm physics, the total cross section for $e^+e^-\to\psi(3770)$ appeared to be $\approx10$~nb, while direct measurements of $D$ meson production indicated that  $e^+e^-\to\psi(3770)\to\bar DD$ was closer to 5~nb. This difference would imply that the $\psi(3770)$ had significant branching fractions to final states without charm, an idea quite at odds with our basic understanding of this state.

This gap was recently closed by CLEO-c who measured~\cite{He:2005bs} $\sigma(e^+e^-\to\psi(3770)\to\bar DD)=6.39\pm0.10^{+0.17}_{-0.08}$~nb, and~\cite{Besson:2005hm} $\sigma(e^+e^-\to\psi(3770)\to{\rm hadrons})=6.38\pm0.08^{+0.41}_{-0.30}$~nb. This puts an upper limit on the ``gap'' of about 10\%. We note that CLEO-c also searched for various other decay modes of the $\psi(3770)$, in several different publications~\cite{Huang:2005fx,Adam:2005mr,Coan:2005ps,Briere:2006ff,Adams:2005ks}, and those observed add up to about 2\%.

Recently, however, BES~\cite{Ablikim:2006aj,BES:2006zq} has reported evidence for a substantially larger gap of $16\pm8$\%. More study will be needed, possibly with BES~III, in order to settle the issue.

\section{NEW STATES IN CHARMONIUM}

\subsection{Discovery of $^1P_1$ Charmonium}

\begin{figure}[ht]
\begin{center}
\includegraphics[width=80mm]{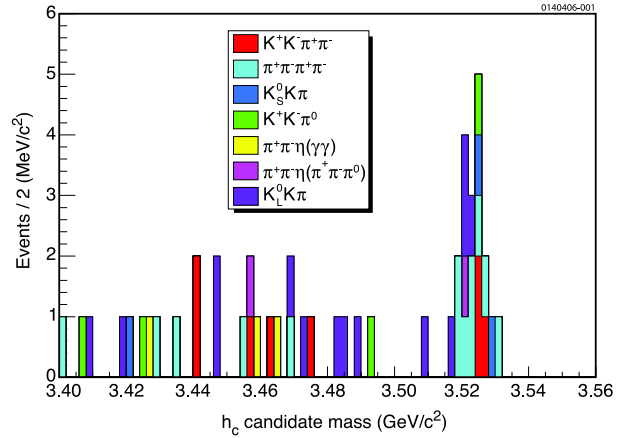}
\caption{Observation of the $h_c(3525)$ from CLEO. Exclusive events of the type $\psi(2S)\to\pi^0h_c\to\pi^0\gamma\eta_c$ are used to reconstruct the $h_c$ with the various indicated decay modes of the $\eta_c$.\label{fig:CLEOhc}}
\end{center}
\end{figure}

The spin triplet $c\bar c$ $P$ state partners of the $J/\psi$ and $\psi(2S)$, called $\chi_{c0,1,2}$, have been known for thirty years. The spin-singlet state, however remained undiscovered. Recently, however, this state, the $h_c(3525)$ has been observed in two separate experiments. Fermilab E835~\cite{Andreotti:2005vu} observed the state in the reaction $\bar p p\to h_c\to\gamma\gamma\gamma$. CLEO~\cite{Rosner:2005ry,Rubin:2005px} was able to observe the state in $\pi^0$ emission from the $\psi(2S)$, both in the $\pi^0$ momentum spectrum, and also in exclusive decays $h_c\to\gamma\eta_c$. The latter is shown in Fig.~\ref{fig:CLEOhc}.
After adding up several $\eta_c$ decay modes, a clear $h_c$ peak is observed with almost no background.

\subsection{Excitations of $^3P_j$ Charmonium}

The Belle collaboration has observed a number of new $c\bar c$ states in various reactions. With masses in the range 3.9 to 4~GeV, these are candidates for radial excitations of the spin-triplet $\chi_{cj}$ sates. These discoveries were discussed in detail by R.\ Waldi in this conference, and the reader is referred to his contribution to these proceedings.

\subsection{The $Y(4260)$}

\begin{figure}[ht]
\begin{center}
\includegraphics[width=40mm]{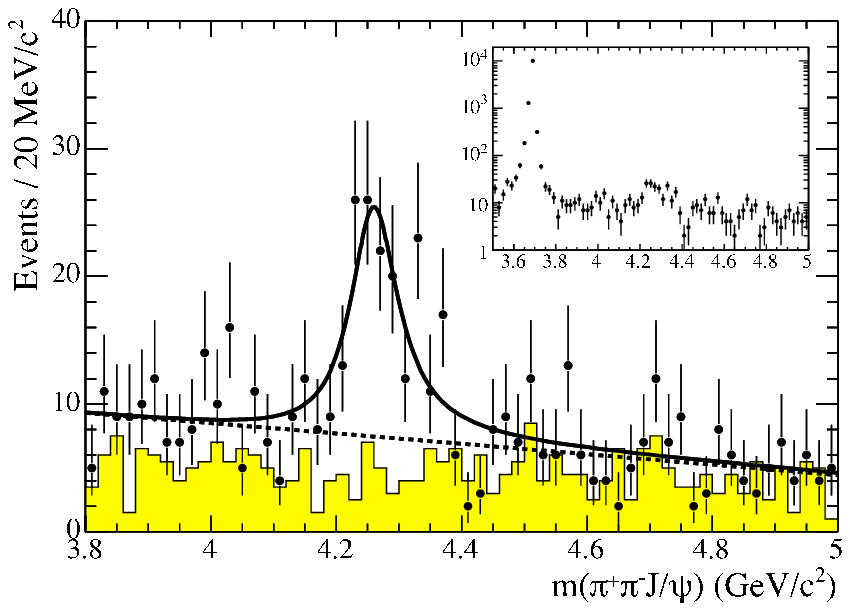}\hfill
\includegraphics[width=40mm]{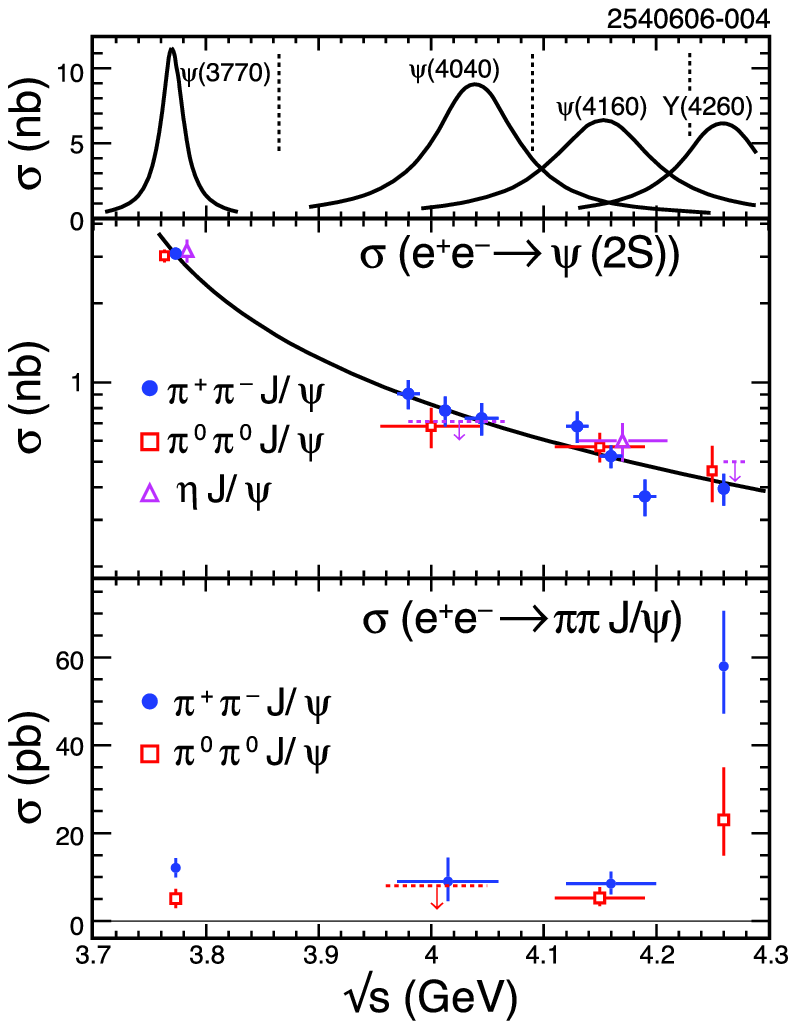}
\caption{Observation of the $Y(4260)$ from BaBar (left) and CLEO (right).\label{fig:Y4260}}
\end{center}
\end{figure}

Babar made a striking discovery~\cite{Aubert:2005rm}, of a new vector charmonium state, the $Y(4260)$. Produced in $e^+e^-$ annihilation via ``radiative return'', that is $e^+e^-\to\pi^+\pi^-J/\psi(\gamma)$, the state is observed via its decay to $J/\psi\pi^+\pi^-$. See Fig.~\ref{fig:Y4260}.
The observation is noteworthy because no vector states are expected at this mass, based on the quark model, and there is no evidence for a peak in the total $e^+e^-$ cross section at this energy.  CLEO-c~\cite{Coan:2006rv} confirmed the BaBar observation (also in Fig.~\ref{fig:Y4260}) using direct $e^+e^-$ annihilation, and also observed the $J/\psi\pi^0\pi^0$ decay mode. No other modes are seen, and the nature of this state remains an enigma. Many theoretical explanations have been put forward, some of which have been addressed by experiment.

\section{CONFRONTING LATTICE QCD}

Recent unquenched calculations in Lattice QCD, making use of ``staggered fermions'', have shown dramatic agreement with experiment~\cite{Davies:2003ik,DeTar:2004tn,Kronfeld:2005fy}, on the order of 1\%. This prescription is not without strong criticism, however~\cite{Neuberger:2004ft}.

Precision QCD calculations have impact on determinations of the CKM matrix and the source of $CP$ violation. Extraction of weak matrix elements typically depends on being able to separate them from hadronic matrix elements, generally in the decays of $B$ mesons. It is a boon to the field if Lattice QCD can determine the strong contributions to the same level of precision as the experimental measurements.

Charm offers an ideal laboratory for testing these predictions. Ultimately it is an experimental question, to determine whether these calculations are indeed valid to the indicated level of precision. The measurements cited by~\cite{Davies:2003ik} existed when the calculations were made. In this section we describe some recent measurements aimed at testing existing calculations.

\subsection{The $D_s$ Decay Constant}

Our first example is the weak decay constant for the $D_s^+$ meson:\\
\centerline{\includegraphics[height=15mm]{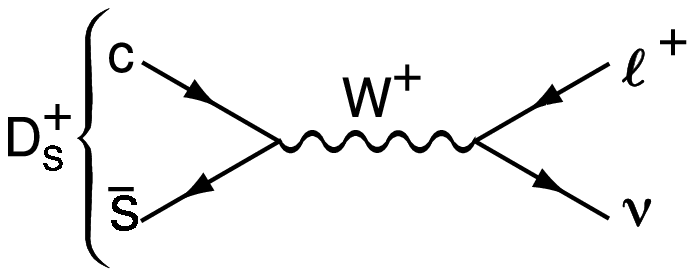}}
The decay constant $f_{D_s}$ is essentially determined by the overlap integral of the charm and strange quark. It is calculable in unquenched Lattice QCD. Using staggered fermion actions, a consortium of the FermiLab Lattice, MILC, and HPQCD collaborations~\cite{Aubin:2005ar} find
$$f_{D_s}=249\pm3\pm16~{\rm MeV}\qquad{\rm (Lattice~QCD)}$$

\begin{figure}[ht]
\begin{center}
\includegraphics[width=45mm]{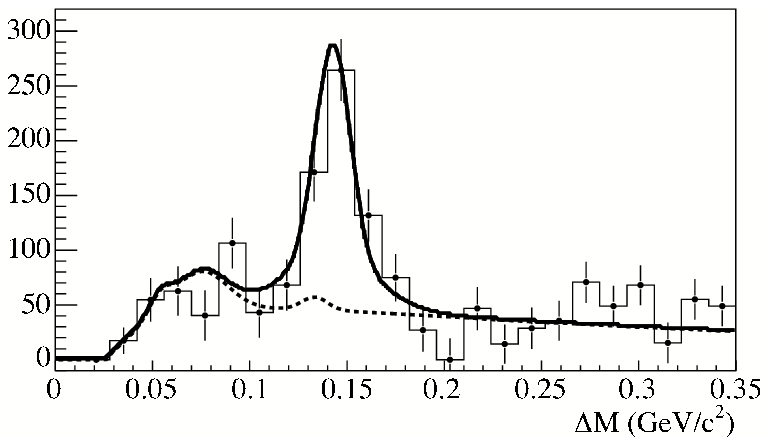}\hfill
\includegraphics[width=35mm]{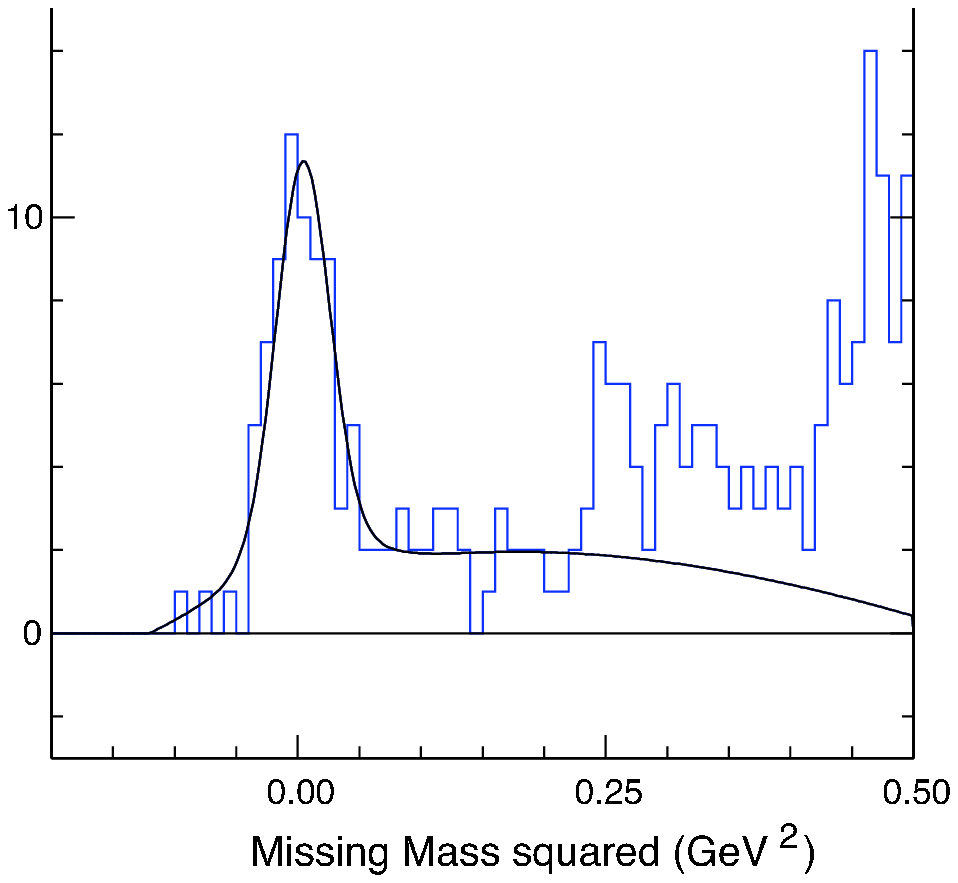}
\caption{Measurement of the $D_s$ decay constant from BaBar (left) and CLEO (right). BaBar observes $D_s^+\to\mu^+\nu_\mu$, and CLEO-c observes both $D_s^+\to\mu^+\nu_\mu$ and $D_s^+\to\tau^+\nu_\tau$.\label{fig:Ds}}
\end{center}
\end{figure}

The decay constant is determined experimentally through a measurement of the decay rate for $D_s^+\to\ell^+\nu$. Figure~\ref{fig:Ds} shows the signals obtained by Babar~\cite{Aubert:2006sd} and CLEO~\cite{Artuso:2006kz}.
The methods are rather different between the two experiments. BaBar produces $D_s$ at high energies and uses a ``charm tagging'' technique to observe $D_s^+\to\mu^+\nu_\mu$. CLEO-c uses the reaction $e^+e^-\to D_sD_s^*$ to kinematically constrain the final state, and observe both $D_s^+\to\mu^+\nu_\mu$ and $D_s^+\to\tau^+\nu_\tau$. They find
$$f_{D_s}=283\pm17\pm7\pm14~{\rm MeV}\qquad{\rm (BaBar)}$$
and
$$f_{D_s}=282\pm16\pm7~{\rm MeV}\qquad{\rm (CLEO,~Preliminary)}$$
The third uncertainty on the BaBar measurement is due to uncertainties in their normalization decay $D_s\to\phi\pi$, which will decrease based on the CLEO-c measurements described in Sec.~\ref{sec:DsHad}.

\subsection{The $D^+$ Decay Constant}

The same test can obviously be performed with $D^+\to\ell^+\nu$, although the experiment is more difficult because this decay is Cabibbo-suppressed. Lattice QCD~\cite{Aubin:2005ar} finds
$$f_{D^+}=201\pm3\pm17~{\rm MeV}\qquad{\rm (Lattice QCD)}$$
With their initial complement of running at the $\psi(3770)$, CLEO-c has determined the branching ratio with precision comparable to the calculation, using a tagging technique made possible by the strongly constrained kinematics. Based on 50 $D^+\to\mu^+\nu_\mu$ events, with an estimated background of 3, CLEO-c~\cite{Artuso:2005ym} finds
$$f_{D^+}=222.6\pm16.7^{+2.8}_{-3.4}~{\rm MeV}\qquad{\rm (CLEO)}$$
CLEO-c will improve on the statistical precision of this measurement when the final data set for $\psi(3770)$ is acquired before Spring~2008.

\subsection{Semileptonic Form Factors}

Similarly precise calculations and measurements can be made of $D$ and $D_s$ form factors in semileptonic decays. Once again, testing the Lattice QCD calculations in the charm system, lends itself to making reliable calculations for $B$ decays and determination of CKM matrix elements.

We focus here on $D^0\to K^-\ell^+\nu$ and $D^0\to\pi^-\ell^+\nu$, for which precise Lattice QCD calculations have been carried out~\cite{Okamoto:2003ur,Aubin:2004ej,Kronfeld:2005fy}. These two decays provide complementary tests. Decays to $K\ell\nu$ are Cabibbo-allowed, so high statistics can be achieved, but the $Q$-value is larger for $\pi\ell\nu$ allowing comparisons up to higher momentum transfer, so shorter distances in the $D$ meson structure. This is particularly true for $\pi e\nu_e$.

\begin{figure}[ht]
\begin{center}
\includegraphics[width=80mm]{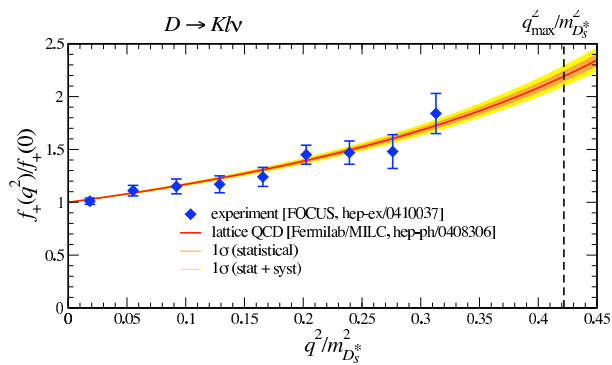}
\caption{Comparison between experiment and calculation for the form factor in $D\to K^-\mu^+\nu_\mu$. The figure is taken from~\cite{Kronfeld:2005fy}.\label{fig:DSemi}}
\end{center}
\end{figure}

Measurements of these form factors have been published by Belle~\cite{Widhalm:2006wz}, BES~\cite{Ablikim:2004ej}, FOCUS~\cite{Link:2004dh}, and CLEO-c~\cite{Poling:2006da}. The most recent comparison between calculation and experiment is shown in Fig.~\ref{fig:DSemi}.
More experimental results from CLEO-c, including $D^0\to\pi^-e^+\nu_e$ are expected soon.

\section{$D$ MIXING AND TESTS OF $CP$}

Neutral meson flavor mixing and $CP$ violation are well known phenomena in the $K$ and $B$ systems. However, unlike $s$ and $b$ quarks, charm quarks are the upper members of their weak doublets, and therefore undergo $CKM$-allowed decays. This has the effect of suppressing mixing and $CP$ violation for neutral $D$ mesons, within the framework of the Standard Model. Conversely, however, this means that searching for these phenomena provides a new window on ``new physics.''

Several recent experimental searches are discussed in this section. We also mention completed and anticipated measurements of charm decays that can be used to better understand mixing and $CP$ violation in the $B$ system.

\subsection{Semileptonic Decay}

The sign of the charged lepton in a semileptonic decay is indicative of the flavor of the decaying meson. That is, $D^0\to X^-\ell^+\nu_\ell$ and $\bar D^0\to X^+\ell^-\bar\nu_\ell$, which are referred to as ``right sign'' (RS) combinations. Leptons with the ``wrong sign'' (WS) combination can only occur through neutral meson mixing. This mixing has a characteristic time signature, so additional information can be gained if the meson decay time can be measured.

Belle~\cite{Abe:2005nq} has measured the ratio of WS to RS events, as a function of $D^0$ decay time. They identify $D^0\to K^{(*)^-}e^+\nu_e$ (and charge conjugate) events through the decay $D^*\to\pi_SD$ where $\pi_S$ is a slow pion. The sign of the $\pi_S$ charge indicates the sign of the $D^*$ and therefore the flavor of the neutral meson. No attempt is made to reconstruct the $K^{*^\pm}$ mesons. The momentum of the missing neutrino is determined by observation of the other particles in the event.

\begin{figure}[ht]
\begin{center}
\includegraphics[width=40mm]{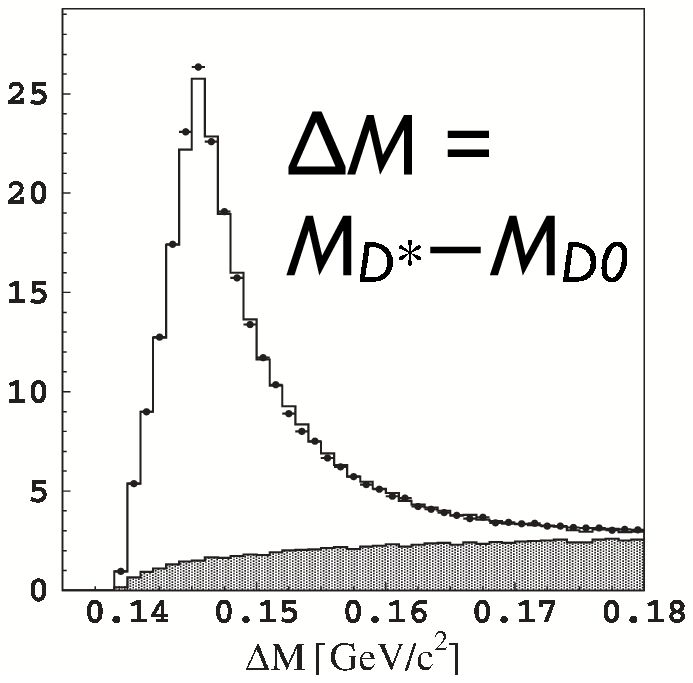}\hfill
\includegraphics[width=40mm]{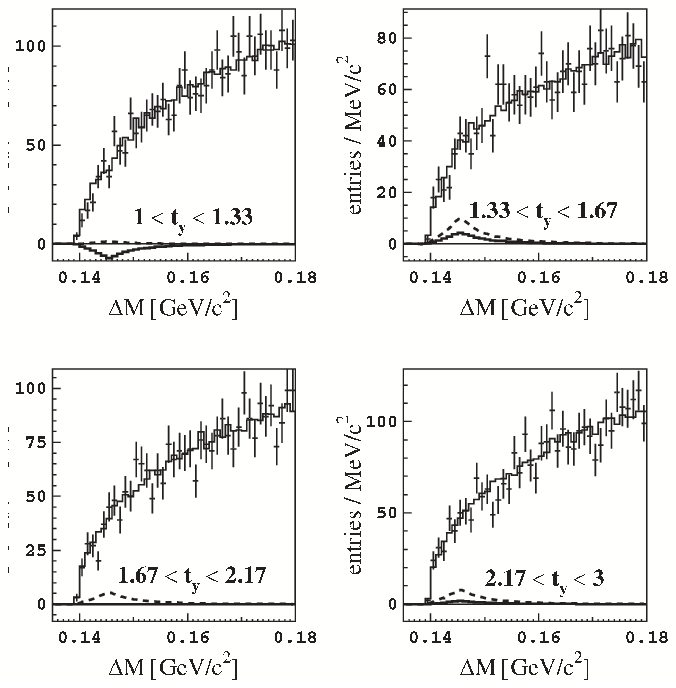}\\
\includegraphics[width=60mm]{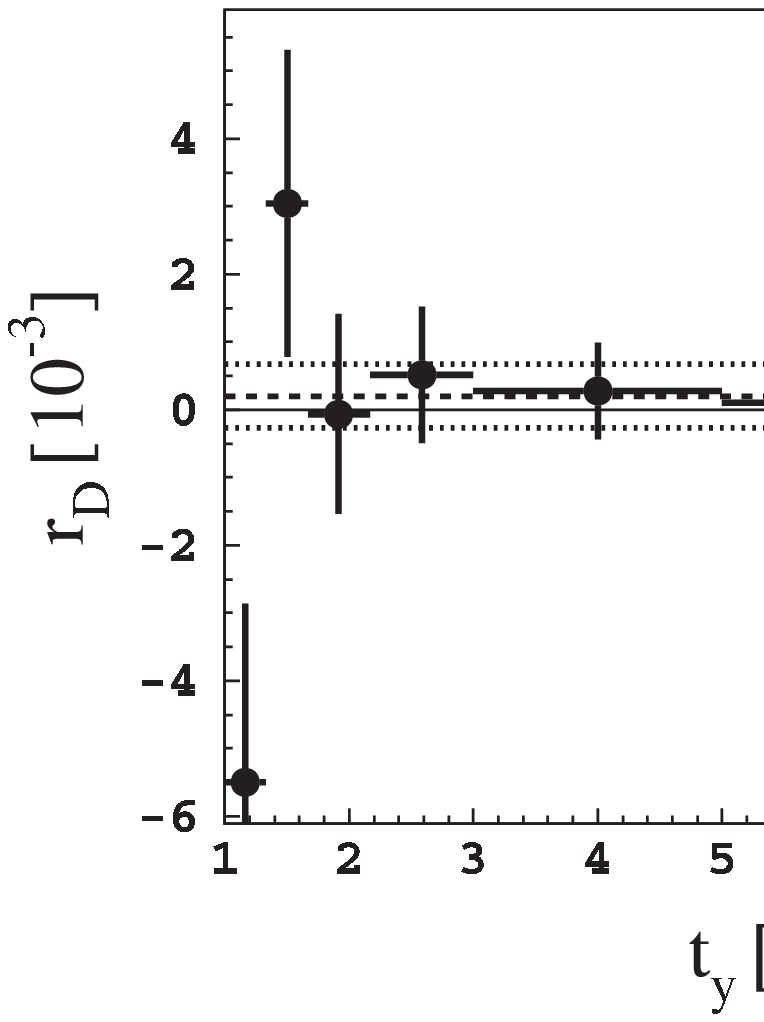}
\caption{Results from Belle~\cite{Abe:2005nq} on $D^0\bar D^0$ mixing using $e^\pm$ semileptonic decays. The upper left shows right-sign (RS) semileptonic signal, reconstructing the $D^*$ and $D$ mass including the missing neutrino. Background is indicated by the shaded histogram. The upper right shows the same quantity, but for wrong-sign (WS) combinations, separated into four decay time bins. None of these plots shows evidence for a signal. The bottom plot shows the ratio $r_D$ between the WS and RW rates, as a function of decay time. An upper limit $r_D<1.0\times10^{-3}$ is derived.\label{fig:BelleMixSL}}
\end{center}
\end{figure}

Their results are shown in Fig.~\ref{fig:BelleMixSL}.
A large signal is evident for the RS combinations. There is no evidence of a signal for the WS events, which are separated into six different decay time bins, four of which are shown. The ratio of WS/RS yields, $r_D$, is also plotted, as a function of decay time. Integrating over time, the experiment determines $r_D<1.0\times10^{-3}$ at 90\% C.L.

\subsection{DCS and Mixing in $D^0\to K^+\pi^-$}

It is also possible to search for mixing in hadronic decays, such as $D^0\to K^+\pi^-$. Of course, such decays may also occur through Double Cabibbo Suppressed (DCS) processes. For single $D^0$ decays, these two effects can only be separated through a measurement of the time dependence.

An excellent illustrative example of this comes from the CDF~II collaboration~\cite{CDF:2006sz}, who have measured the DCS rate $R_D$, relative to the Cabibbo-allowed decay rate $D^0\to K^-\pi^+$. From a $\int{\cal L}dt=0.35~{\rm fb}^{-1}$ sample of $\bar pp$ collisions at $\sqrt{s}=1.96$~TeV at the FermiLab Tevatron, they extract $2005\pm104$ DCS events, yielding $R_D=0.405\pm0.021\pm0.011\%$.

\begin{figure}[ht]
\begin{center}
\includegraphics[height=38mm]{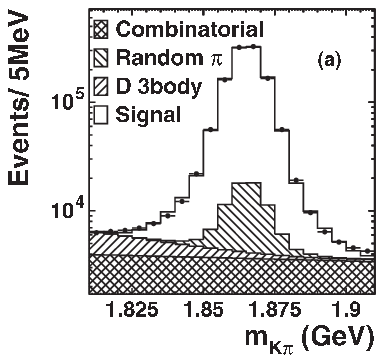}\hfill
\includegraphics[height=38mm]{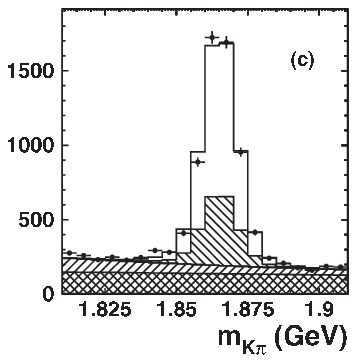}\\
\includegraphics[width=80mm]{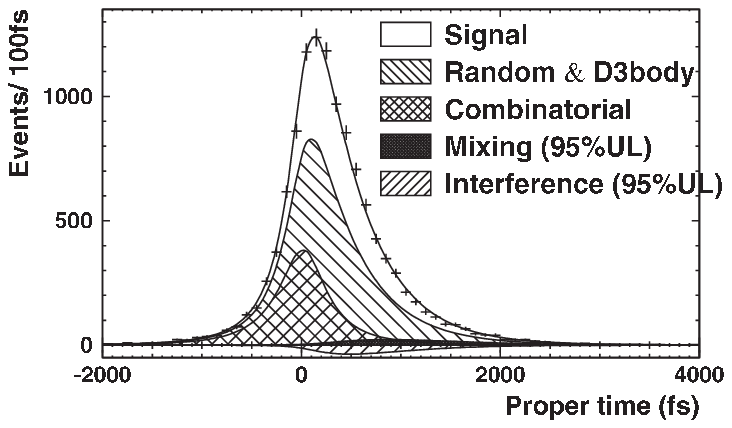}
\caption{Results from Belle~\cite{Zhang:2006dp} on $D^0\bar D^0$ mixing using $K\pi$ hadronic decays. The upper plots show the $K\pi$ invariant mass for right-sign (left) and wrong-sign (right) combinations. The various backgrounds are indicated. The bottom plot shows the distribution of the WS events in decay time. The Doubly Cabibbo Suppressed (DCS) decay rate is found to be $R_D=0.377\pm0.008\pm0.005\%$, relative to that for Cabibbo Allowed decay. There is no evidence for mixing from the time distribution.\label{fig:BelleMixHad}}
\end{center}
\end{figure}

Belle~\cite{Zhang:2006dp} has carried this work further, including a measurement of $R_D$ as a function of decay time. Their results are shown in Fig.~\ref{fig:BelleMixHad}.
The RS and WS $K\pi$ combinations are plotted, and various backgrounds are indicated. The WS combinations are also plotted as a function of time, and the 95\%~C.L. upper limit is indicated, based on the expected time distribution. Integrating over time, Belle finds an improved WS/RS ratio of $R_D=0.377\pm0.008\pm0.005\%$.

\subsection{Summary}

\begin{figure}[ht]
\begin{center}
\includegraphics[width=40mm]{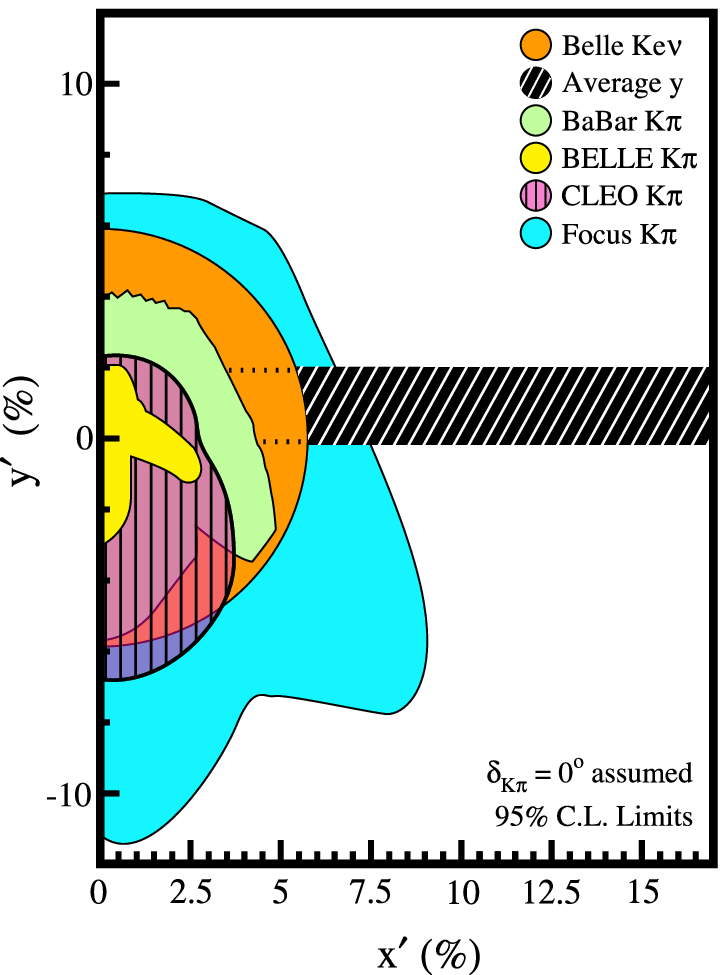}\hfill
\includegraphics[width=40mm]{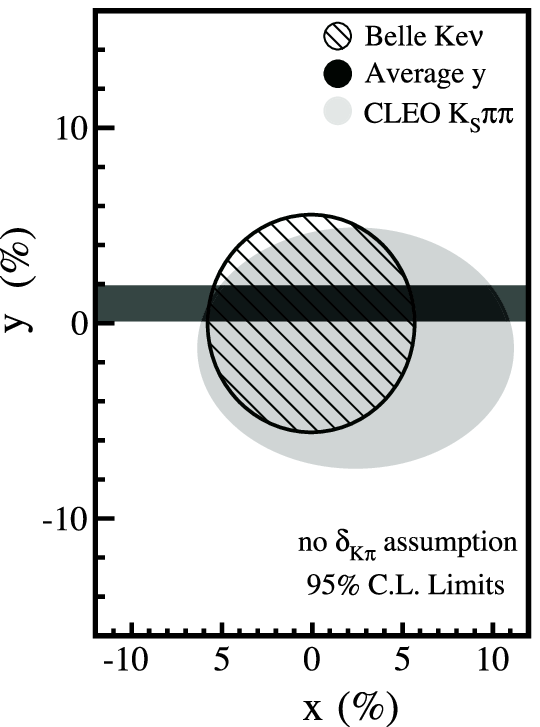}
\caption{Summary of $D^0\bar D^0$ mixing searches from the Particle Data Group~\cite{PDBook06}.\label{fig:DMixSummary}}
\end{center}
\end{figure}

David Asner has reviewed $D^0\bar D^0$ mixing in the most recent Particle Data Group compilation~\cite{PDBook06}. The results from various experiments are shown in Fig.~\ref{fig:DMixSummary}.
In these plots, $x\equiv(M_1-M_2)/\Gamma$, $y\equiv(\Gamma_1-\Gamma_2)/2\Gamma$, where $M_{1,2}$ and $\Gamma_{1,2}$ are the masses and widths of the $CP$ eigenstate combinations of $D^0$ and $\bar D^0$, and $\Gamma\equiv(\Gamma_1+\Gamma_2)/2$. Also $x^\prime$ and $y^\prime$ are rotated relative to $x$ and $y$ by a ``strong phase difference'' between Doubly Cabibbo Suppressed and Cabibbo Favored processes not attributed to the first order electroweak spectator diagram.

No experiments show evidence for $D^0\bar D^0$ mixing. These results are consistent with the standard model, and put constraints on some models of ``new physics."

It should be noted that analyses of $D^0\bar D^0$ generally assume that $CP$ is conserved. Indeed, one expects this effect to be even smaller than mixing. Nevertheless, specific searches for $CP$ violation in neutral $D$ decay have been carried out. For example, see the review by Kirkby and Nir in~\cite{PDBook06}. $CP$ violation could be observed, in principle, from differences in the decay rates, or phase space distributions, between particles and antiparticles. It might also be detected through $T$ violation, albeit indirectly by invoking the $CPT$ theorem.

At this conference, the presenter presented a summary table of results taken from a talk by S.\ Stone at FPCP06~\cite{Stone:2006bb}. Table~\ref{tab:StoneCPV} reproduces this summary.
\begin{table}[ht]
\begin{center}
\caption{Recent measurements of $CP$ or $T$ violating asymmetries in charm decays.}
\begin{tabular}{|l|l|c|}
\hline \textbf{Expt} & \textbf{Decay Mode} & \textbf{$A_{CP}$ or $A_T$ (\%)}\\
\hline
BaBar~\cite{Aubert:2005gj} & $D^+\to K^-K^+\pi^+$ & $1.4\pm1.0\pm0.8$\\
\hline
BaBar~\cite{Aubert:2005gj} & $D^+\to\phi\pi^+$ & $0.2\pm1.5\pm0.6$\\
\hline
BaBar~\cite{Aubert:2005gj} & $D^+\to K^{*^0}K^+$ & $0.2\pm1.5\pm0.6$\\
\hline
Belle~\cite{Tian:2005ik} & $D^0\to K^+\pi^-\pi^0$ & $-0.6\pm5.3$\\
\hline
Belle~\cite{Tian:2005ik} & $D^0\to K^+\pi^-\pi^+\pi^-$ & $-1.8\pm4.4$\\
\hline
CLEO~\cite{Cronin-Hennessy:2005sy} & $D^0\to\pi^+\pi^-\pi^0$ & $1^{+9}_{-7}\pm8$\\
\hline
CLEO~\cite{Kopp:2000gv} & $D^0\to K^-\pi^+\pi^0$ & $-3.1\pm8.6$\\
\hline
CLEO~\cite{Asner:2003uz} & $D^0\to K_S\pi^+\pi^-$ & $-0.9\pm2.1^{+1.0+1.3}_{-4.3-3.7}$\\
\hline
CDF~\cite{Acosta:2004ts} & $D^0\to K^+K^-$ & $2.0\pm1.2\pm0.6$\\
\hline
CDF~\cite{Acosta:2004ts} & $D^0\to\pi^+\pi^-$ & $1.0\pm1.3\pm0.6$\\
\hline
FOCUS~\cite{Link:2005th} & $D^0\to K^+K^-\pi^+\pi^-$ & $1.0\pm5.7\pm3.7$\\
\hline
FOCUS~\cite{Link:2005th} & $D^+\to K^0K^-\pi^+\pi^-$ & $2.3\pm6.2\pm2.2$\\
\hline
FOCUS~\cite{Link:2005th} & $D_s^+\to K^0K^-\pi^+\pi^-$ & $-3.6\pm6.7\pm2.3$\\
\hline
\end{tabular}
\label{tab:StoneCPV}
\end{center}
\end{table}
As listed in this table, BaBar compares $D^+$ and $D^-$ decay rates; CLEO compares phase space distributions through a Dalitz plot analysis; CDF measures decay rates for $D^0$ and $\bar D^0$ using $D^{*^+}\to\pi^+D^0$ for a flavor tag; and the FOCUS results make use of triple product correlations to search for $T$ violation. In no case is there significant evidence for $CP$ or $T$ violation.

\subsection{An Aside: $D^0\to K^*K$}

The $b\to cW^-$ transition is the least-forbidden weak decay mode for the $b$ quark, so the dynamics of charm decay can be a window on $b$ decay. For example, the amplitudes for $B^-\to K^-D^0$ and $B^-\to K^-\bar D^0$ are about the same size in magnitude, but differ by a phase which can be used to study $CP$ violation in $B^-$ decay. Understanding the decay modes of the neutral $D$ mesons is key to unraveling this physics.

\begin{figure}[ht]
\begin{center}
\includegraphics[width=80mm]{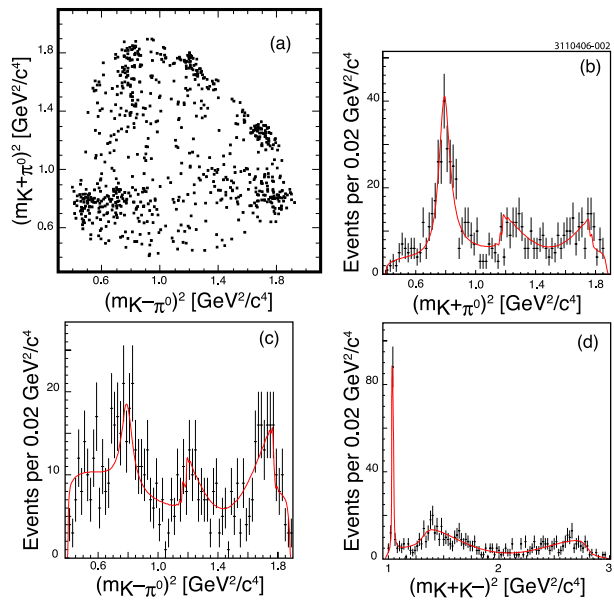}
\caption{Dalitz plot, and projections including the multi-resonance fit, for the decay $D^0\to K^+K^-\pi^0$ from CLEO-III~\cite{ParasKKPi}. The behavior of the $K^{*^+}$ and $K^{*^-}$ bands for low $K\pi$ mass indicated destructive interference with each other, through their interference with the underlying $K^+K^-$ $S$-wave.\label{fig:Paras}}
\end{center}
\end{figure}

CLEO-III recently published a measurement~\cite{ParasKKPi} of the interference between $K^{*+}K^-$ and $K^{*-}K^+$ amplitudes in the decay $D^0\to K^+K^-\pi^0$. This is a necessary result in order to extract the CKM angle $\gamma$ (aka $\phi_3$) in $B\to KD$ decay. By fitting the Dalitz plot structure for $D^0\to K^+K^-\pi^0$, the authors are able to determine the ``strong phase'' difference between the $K^{*+}K^-$ and $K^{*-}K^+$ modes. Their data and fit results are shown in Fig.~\ref{fig:Paras}.
The find nearly total destructive interference between the two $K^*$ amplitudes, leading to a strong phase difference $\delta_D=332^\circ\pm8^\circ\pm11^\circ$. They also find a value of $0.52\pm0.05\pm0.04$ for the ratio of the magnitudes of the $K^{*-}K^+$ amplitude to the $K^{*+}K^-$ amplitude. 

\subsection{Quantum Correlations}

Production of $D^0\bar D^0$ pairs through the reaction $e^+e^-\to\psi(3770)\to D^0\bar D^0$ provides unique opportunities to study mixing and $CP$ violation. Since the two mesons are produced in a pure quantum state, ``wrong'' admixtures of flavor or $CP$ in one decay become apparent linearly through interference with the other meson. This is in contrast to searches using decay rates directly, in which case these small amplitudes are realized through their square.

Asner and Sun~\cite{Asner:2005wf} have worked out the formalism for these ``quantum correlated'' observations of neutral $D$ mesons, and have estimated the sensitivity one expects to achieve with the anticipated data set from CLEO-c. Cinabro~\cite{Cinabro:2006qv} and Sun~\cite{Sun:2006gn} have discussed this as well, with some reference to preliminary data from CLEO-c.

\begin{figure}[ht]
\begin{center}
\includegraphics[angle=-90,width=80mm]{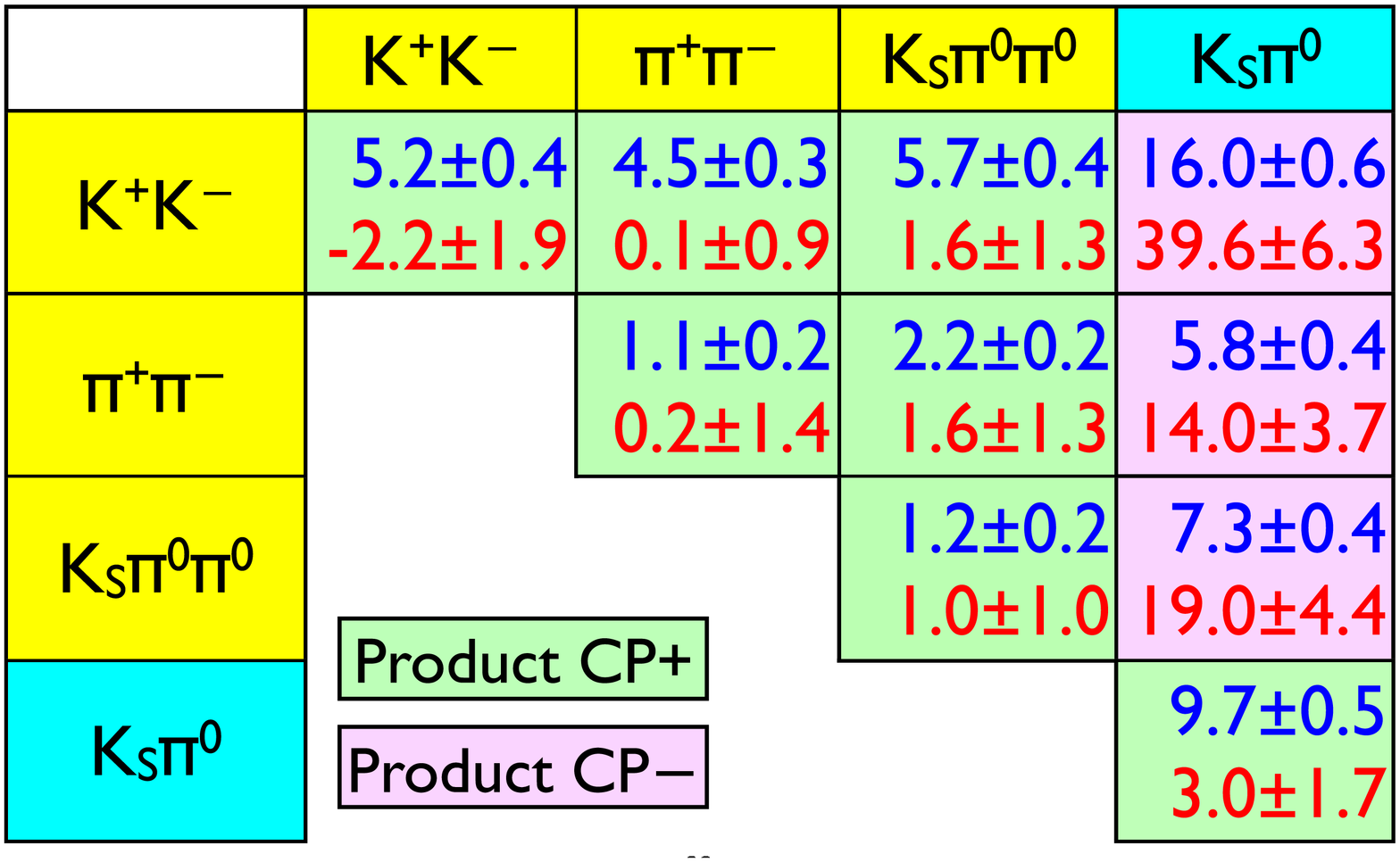}
\caption{Preliminary results on measurements of quantum correlated neutral $D$ mesons, from CLEO-c.\label{fig:QuCorr}}
\end{center}
\end{figure}

The presenter showed some of these preliminary data, which are reproduced in Fig.~\ref{fig:QuCorr}.
All values are numbers of events where both neutral $D$ mesons are observed. The upper value in each cell is from a Monte Carlo simulation, and the lower value is from CLEO-c data. The simulation is normalized to the same luminosity as the data. Known branching ratios are used in the simulation, but no quantum correlation is included. The four decay modes considered are $K^+K^-$, $\pi^+\pi^-$, $K_S\pi^0\pi^0$ (all $CP=+1$) and $K_S\pi^0$ ($CP=-1$). One observes that for decay combinations with product $CP$ negative, the number of observed events is twice that for the Monte Carlo. On the other hand, if the produce $CP$ is positive, the number of observed events is consistent with zero. This is expected, since although the $CP$ of the $\psi(3770)$ is $+1$, the $D^0\bar D^0$ pair is produced in a relative $P$-state.

\section{THE FUTURE}

Expect continued high precision measurements from the Belle, BaBar, CDF, and D0 collaborations. These experiments produce large amounts of charm, and their analysis procedures are well oiled.

CLEO-c will run through March 2008, including an increased data set at the $\psi(3770)$. This data set will be thoroughly exploited for quantum correlations, precision charm measurements, and other physics.

BES~III will be coming on line in the next few years. The BEPC~II storage ring system should deliver extremely high luminosity for $\sqrt{s}$ near 4~GeV. It is not unreasonable to expect $\approx25\times$ as many $D\bar D$ events as will be available from CLEO-c.

Finally, LHCb and PANDA will begin to take data over the next several years. Charm has a bright future.

\bigskip
\begin{acknowledgments}
I am very grateful to Joao de Mello and his colleagues for organizing a splendid conference. Many thanks to colleagues on other experiments for making their results available on a preliminary basis. Thanks also to my CLEO collaborators, especially Roy Briere who provided invaluable help in assembling the latest results in charm physics.
\end{acknowledgments}

\bigskip

\bibliographystyle{unsrt}
\bibliography{CharmBib}

\end{document}